\documentclass[%
 aip,
 amsmath,amssymb,
 reprint,%
]{revtex4-1}

\usepackage{graphicx}
\usepackage{dcolumn}
\usepackage{bm}

\usepackage[utf8]{inputenc}
\usepackage[T1]{fontenc}
\usepackage{mathptmx}
\usepackage{etoolbox}
\usepackage{adjustbox}
\usepackage{hyperref}
\usepackage{cleveref}
\DeclareUnicodeCharacter{2212}{-}

\makeatletter
\def\@email#1#2{%
 \endgroup
 \patchcmd{\titleblock@produce}
  {\frontmatter@RRAPformat}
  {\frontmatter@RRAPformat{\produce@RRAP{*#1\href{mailto:#2}{#2}}}\frontmatter@RRAPformat}
  {}{}
}%
\makeatother
\begin{document}

\preprint{AIP/123-QED}

\title{Exploring the Potential of Residual Impurities in Germanium Detectors for MeV-Scale Dark Matter Detection}
\author{Dongming Mei}%
 \email{dongming.mei@usd.edu}
$\affiliation{ 
University of South Dakota}$

\date{\today}

\begin{abstract}
The direct detection of MeV-scale dark matter (DM) particles hinges on achieving an exceptionally low energy detection threshold. Germanium (Ge) detectors, meticulously tailored with precise impurity compositions, hold the potential to enhance sensitivity to energy levels below the sub-electronvolt (sub-eV) range. This study explores the behavior of residual impurities inherent to Ge detectors at helium temperatures, unveiling a captivating freeze-out phenomenon leading to the formation of excited localized states known as dipole states. Using compelling evidence from relative capacitance measurements obtained from two detectors, we elucidate the transition of impurity atoms from free charge states to these dipole states as the temperature drops from 11 K to 6.5 K. Our investigation comprehensively covers the intricate formation of these dipole states in both n-type and p-type impurities. Furthermore, we shed light on the electric field generated by these dipole states, revealing their ability to trap charges and facilitate the creation of cluster dipole states. Confirming findings from previous measurements, we establish that these excited dipole states exhibit a binding energy of less than 10 meV, offering an exceptionally low detection threshold for MeV-scale DM. Building upon this concept, we propose the development of a 1-kg Ge detector with internal charge amplification—an innovative approach poised to surpass electrical noise and enable the detection of MeV-scale DM with unprecedented sensitivity.

\end{abstract}
\maketitle
\section{Introduction}
 
 MeV-scale dark matter (DM) particles have emerged as prominent candidates for DM~\cite{ess2012, ess2016, ho, ste}. Direct searches for MeV-scale DM, involving energy deposition resulting from DM coupling to ordinary matter, represent a cutting-edge frontier in particle physics and astrophysics. Despite recent efforts and advances, significant challenges and limitations persist, primarily due to a detection energy threshold higher than $\sim$100 eV, rendering a substantial parameter space inaccessible with current technology.

A primary challenge in MeV-scale DM searches is the need for detectors with ultra-low energy thresholds, capable of measuring the minute energy depositions resulting from interactions between MeV-scale DM and detector materials~\cite{ess2012, mei}. Achieving such low detection thresholds demands extraordinary sensitivity, often requiring detectors capable of detecting single electron-hole (e-h) pairs~\cite{ess2012, mei}.

The central challenges primarily revolve around electronic noise present at a level of $\sim$100 eV, hindering the detection of small energy signals produced by MeV-scale DM interactions. This noise originates from intricately designed components within the data acquisition system, operating within the 100 eV range, and poses a significant background concern that requires meticulous consideration~\cite{mei}. An ongoing challenge persists, focusing on attaining essential detection energy thresholds while effectively mitigating electronic noise~\cite{mei}.

To overcome these limitations, researchers have proposed innovative techniques like Germanium Internal Charge Amplification (GeICA)\cite{geia}. GeICA significantly amplifies the charge signal, effectively surpassing electronic noise and potentially reducing the detection energy threshold to around 0.1 eV~\cite{mei}. This method relies on impact ionization, a phenomenon observed in Ge diodes~\cite{imp1, imp}. The successful implementation of GeICA depends on generating the necessary electric fields to trigger impact ionization within the detector material~\cite{mei}.

In 2018, Mei et al.\cite{mei} introduced the concept of GeICA as a novel approach for detecting MeV-scale DM through the impact ionization of impurities\cite{geia, imp1, imp}. This necessitates operating a Ge detector at a temperature range where the depletion process becomes superfluous, resulting in the absence of free charges within the detector. This paper delves into a comprehensive study of Ge's properties as a function of temperature and makes a remarkable discovery: when the temperature drops below 6.5 Kelvin (K), impurity atoms effectively freeze out. An intriguing consequence of this freeze-out is the constancy observed in the relative capacitance of two Ge detectors below 6.5 K, matching the detector capacitance at nitrogen temperatures when the device was previously depleted.  This paper provides a thorough physical explanation for impurity freeze-out and the formation of dipole states, shedding light on the generation of electric fields by these dipole states. Furthermore, it elucidates how drift charges can be captured to create cluster dipole states. The paper concludes by introducing an optimized detector design, strategically engineered to achieve an amplification factor surpassing 100. This design adopts a planar detector geometry, featuring carefully selected impurity levels, and operates at helium temperatures, specifically tailored for MeV-scale DM searches.

\section{Impurity Freeze-out Concept}
In this section, we initially presented data detailing the capacitance measurements of two Ge detectors—namely, one n-type and one p-type—in a controlled laboratory setting, plotted against varying temperatures. Subsequently, we demonstrated a clear correlation between the observed sharp decline in capacitance below 11 K and the concurrent freeze-out of free charges, coupled with the formation of dipoles involving both free charges and impurities.

\begin{figure} [htbp]
  \centering
  \includegraphics[clip,width=0.9\linewidth]{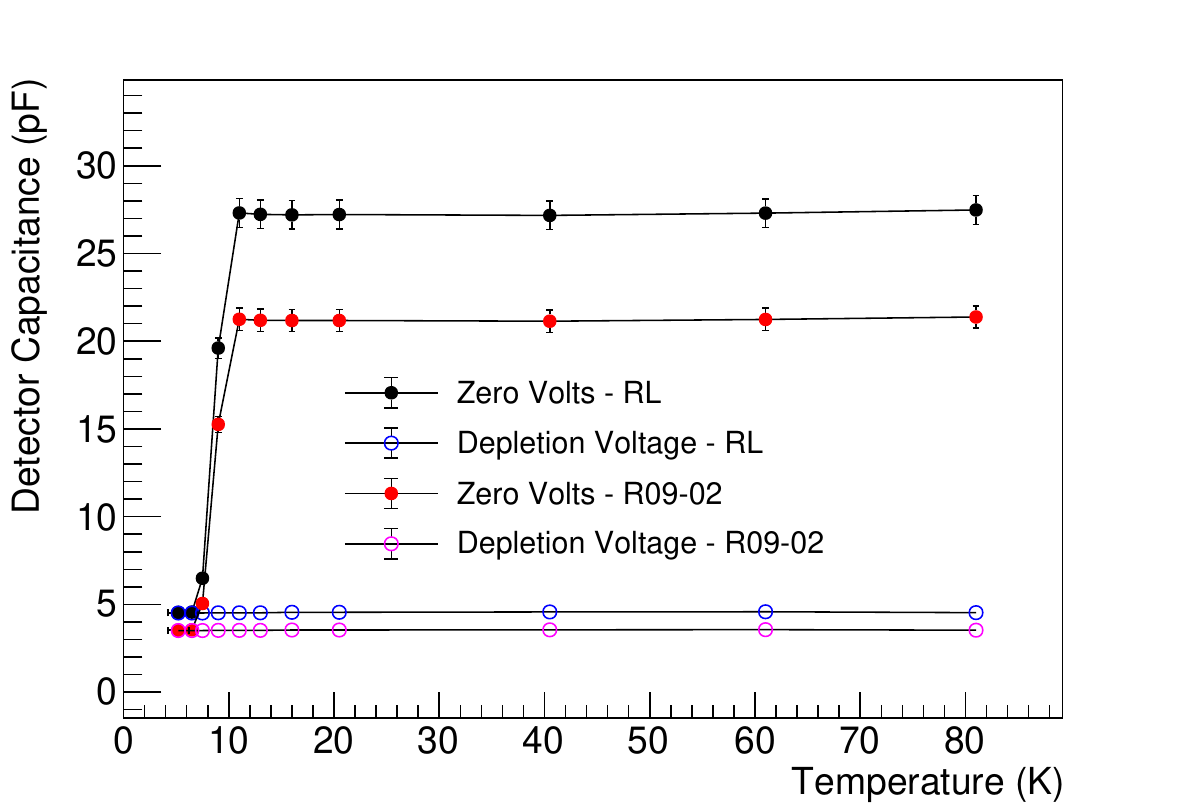}
  \caption{The data points illustrate the capacitance (R09-02 and RL) as a function of temperature under two conditions: one with zero bias voltage and the other with the depletion voltage applied. R09-02 is subjected to a depletion voltage of 1200 volts, while RL undergoes a depletion voltage of 400 volts. The error in capacitance is within 3\%, and the temperature error is within 0.5 K.}
  \label{fig:freeze-out1}
\end{figure}

Two detectors, namely an n-type (R09-02) and a p-type (RL) with corresponding net impurity densities of 7.02$\times$10$^{10}$/cm$^{3}$ and 6.2$\times$ 10$^{^9}$/cm$^{3}$, were cooled down from 80 K to 5.2 K~\cite{mei2022evidence}. The dimensions of R09-02 are 11.7 × 11.5 × 5.5 mm$^3$, while RL has dimensions of 18.8 × 17.9 × 10.7 mm$^3$. The relative capacitance of these planar Ge detectors was measured at various temperatures, initially without bias voltage and subsequently at the depletion voltage, occurring when the detector is operated at $\sim$80 K~\cite{mei2022evidence}. The relative capacitance is defined as the capacitance with zero bias voltage relative to the capacitance at the same temperature when the detectors were fully depleted. The depletion voltage for R09-02 is 1200 volts, while the depletion voltage for RL is 400 volts. The capacitance of two detectors, when both were depleted around 80 K, are 3.51 pF for R09-02 and 4.51 pF for RL, respectively. Figures~\ref{fig:freeze-out1} depicts the capacitance versus temperature when normalized to the capacitance of both detectors at the depletion voltage around 80 K.

As depicted in Figures~\ref{fig:freeze-out1}, it is evident that the capacitance of both detectors remains nearly constant from 80 K down to approximately 11 K under zero bias voltage. However, as the temperature decreases from 11 K to 6.5 K, the relative capacitance experiences a rapid decline, indicative of a substantial temperature gradient. Subsequently, as the temperature continues to decrease from 6.5 K to 5.2 K, the capacitance stabilizes at a consistent level. Remarkably, this level matches the relative capacitance observed when the detector is depleted at 80 K and subsequently cooled down to 5.2 K, as illustrated in Figures~\ref{fig:freeze-out1}.

This observation indicates the absence of free charges within the detector volume when the Ge detector is cooled down to 6.5 K. This is a consequence of the impurities in the Ge detector transitioning into localized states as temperatures drop below 11 K~\cite{mei2022evidence}. This phenomenon is commonly referred to as "freeze-out." In this state of freeze-out, the proportion of electrons bound to donors in an n-type material with an impurity density of N$_{d}$ can be expressed as:
\begin{equation}
    \frac{n_d}{n+n_d} = \frac{1}{\frac{N_C}{2N_d}exp[-\frac{(E_C-E_d)}{k_BT}]+1},
\end{equation}
where $n_d$ corresponds to the number of electrons bound to donors, $n$ represents the total free electrons in the conduction band, $N_C$ signifies the density of states in the conduction band, $E_C$ denotes the energy level of the conduction band, and $E_d$ signifies the energy level of donor impurity. The Boltzmann constant is denoted as $k_{B}$, and $T$ represents temperature.  The factor of $\frac{1}{2}$ is attributed to the existence of two available states at a donor site, corresponding to the two distinct spin states of an electron.
A similar results can be produced for a p-type material with impurity density of N$_a$:
\begin{equation}
    \frac{p_a}{p+p_a} = \frac{1}{\frac{N_V}{4N_a}exp[-\frac{(E_a-E_V)}{k_BT}]+1},
\end{equation}
where $p_a$ corresponds to the number of holes bound to acceptors, $p$ represents the total free holes in the valence band, $N_V$ signifies the density of states in the valence band, $E_{a}$ denotes the energy level of the acceptor impurity, and $E_{V}$ represents the energy level of the valence band. The factor of $\frac{1}{4}$ is attributed to the combination of two bands (light hole and heavy hole) and the existence of two spin states.

\begin{figure} [htbp]
  \centering
  \includegraphics[clip,width=1.0\linewidth]{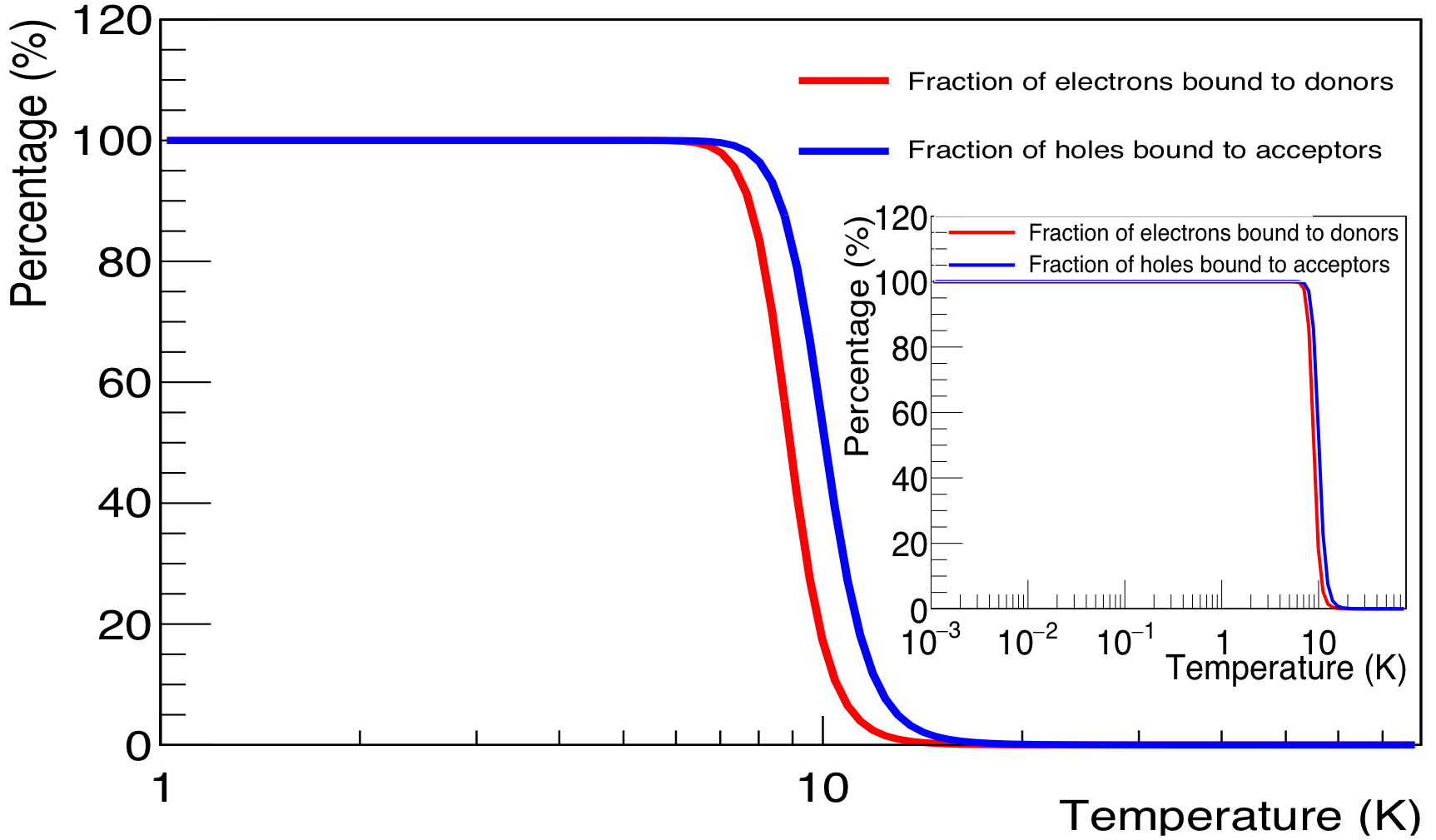}
  \caption{Displayed is the temperature-dependent ratio of electrons or holes bound to donors or acceptors. N$_{d}$ = 7.02$\times$10$^{10}$/cm$^3$ and E$_{C}$ - E$_{d}$ = 0.01 eV are used in generating this plot. Similarly, N$_{a}$ = 6.2$\times$10$^{9}$/cm$^3$ and E$_{a}$ - E$_{V}$ = 0.01 eV are applied. Please note that the inset plot depicts the same information over a broader temperature range. }
  \label{fig:bound}
\end{figure}

The proportion of electrons bound to donors or the fraction of holes bound to acceptors is illustrated in Figure~\ref{fig:bound}.

\begin{figure} [htbp]
  \centering
  \includegraphics[clip,width=0.9\linewidth]{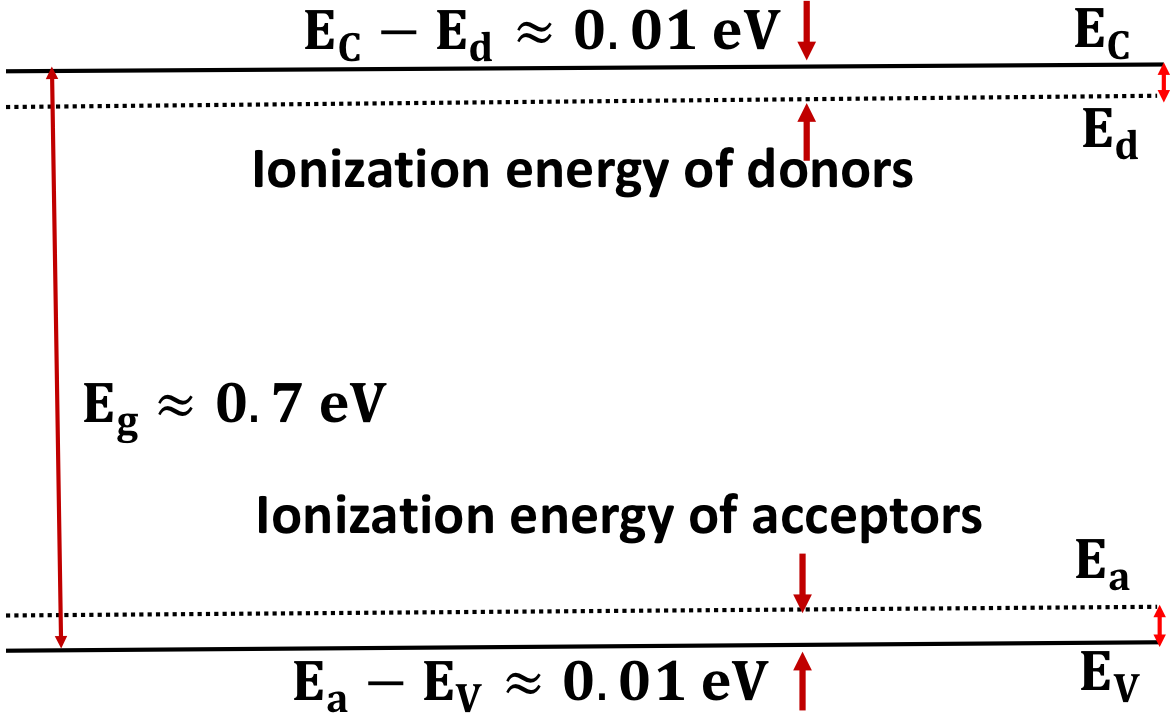}
  \caption{Shown is a sketch of the band structure of Ge featuring the presence of residual impurities.}
  \label{fig:band}
\end{figure}

When we compare Figure~\ref{fig:bound} to Figures~\ref{fig:freeze-out1}, it becomes clear that the transition occurs within a remarkably narrow temperature range, spanning from 11K to 6.5K in both cases. In Figures~\ref{fig:freeze-out1}, the capacitance exhibits a steep decline from 11K to 6.5K, while the proportion of electrons bound to donors or holes bound to acceptors undergoes a dramatic increase over the same temperature interval. This consistent observation confirms that capacitance is directly related to the overall availability of free charges within the detector volume. It's worth noting that residual impurity atoms remain ionized from 80 K down to about 11 K. However, from 11 K to 6.5 K, these impurity atoms begin to undergo a freeze-out process, transforming into bound states that effectively appear neutral in charge. This rapid transition within the narrow temperature range can be elucidated by referencing the band structures illustrated in Figure~\ref{fig:band} where E$_{C}$ - E$_{d}$ = 0.01 eV and E$_{a}$ - E$_{V}$ = 0.01 eV.

As illustrated in Figure~\ref{fig:bound}, when temperatures fall below 6.5 K, almost 100\% of electrons become bound to donors, and nearly all holes become bound to acceptors. As a result of the binding of all free electrons with donors and all free holes with acceptors, dipole states are generated. These dipole states correspond to excited impurity atoms. Using two examples, one n-type and one p-type, we illustrate the stable formation of dipole states when temperatures drop below 6.5 K.

\begin{figure} [htbp]
  \centering
  \includegraphics[clip,width=0.9\linewidth]{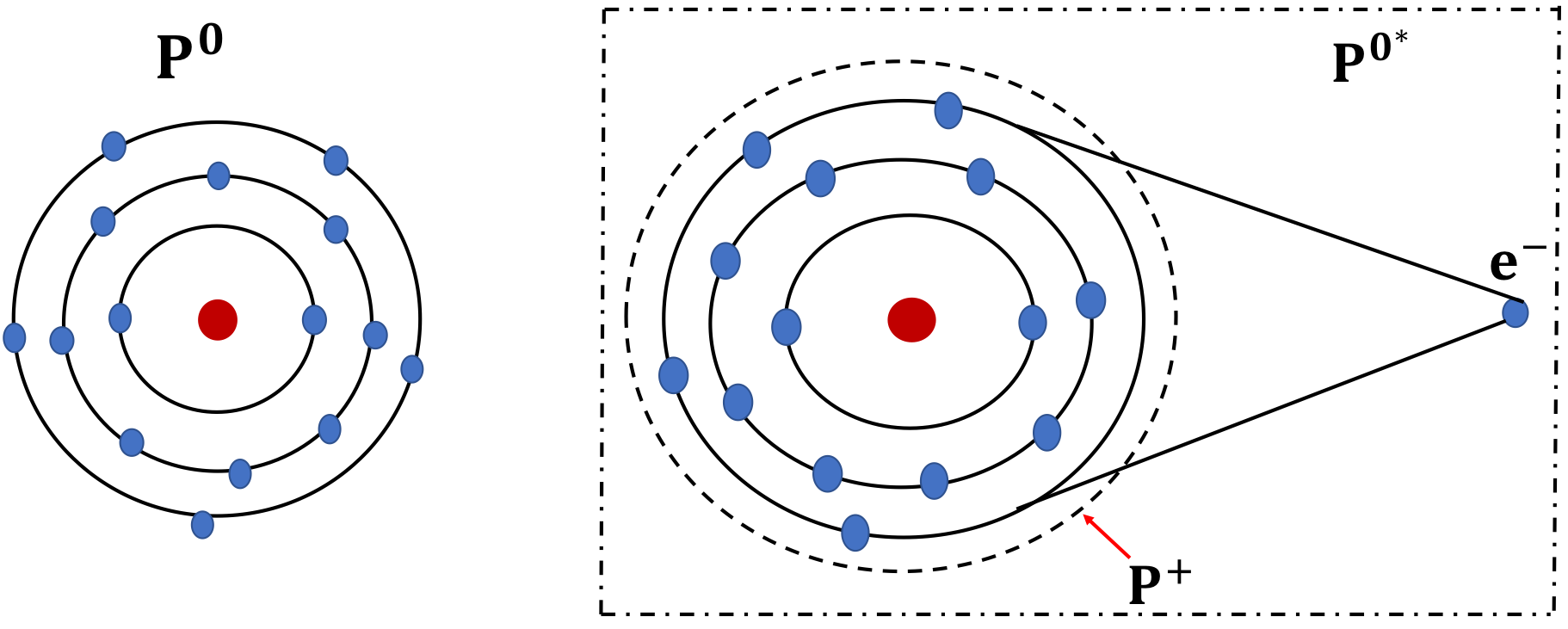}
  \caption{Displayed are the ground state of phosphorus and a phosphorus dipole state.}
  \label{fig:p}
\end{figure}

\begin{figure} [htbp]
  \centering
  \includegraphics[clip,width=0.9\linewidth]{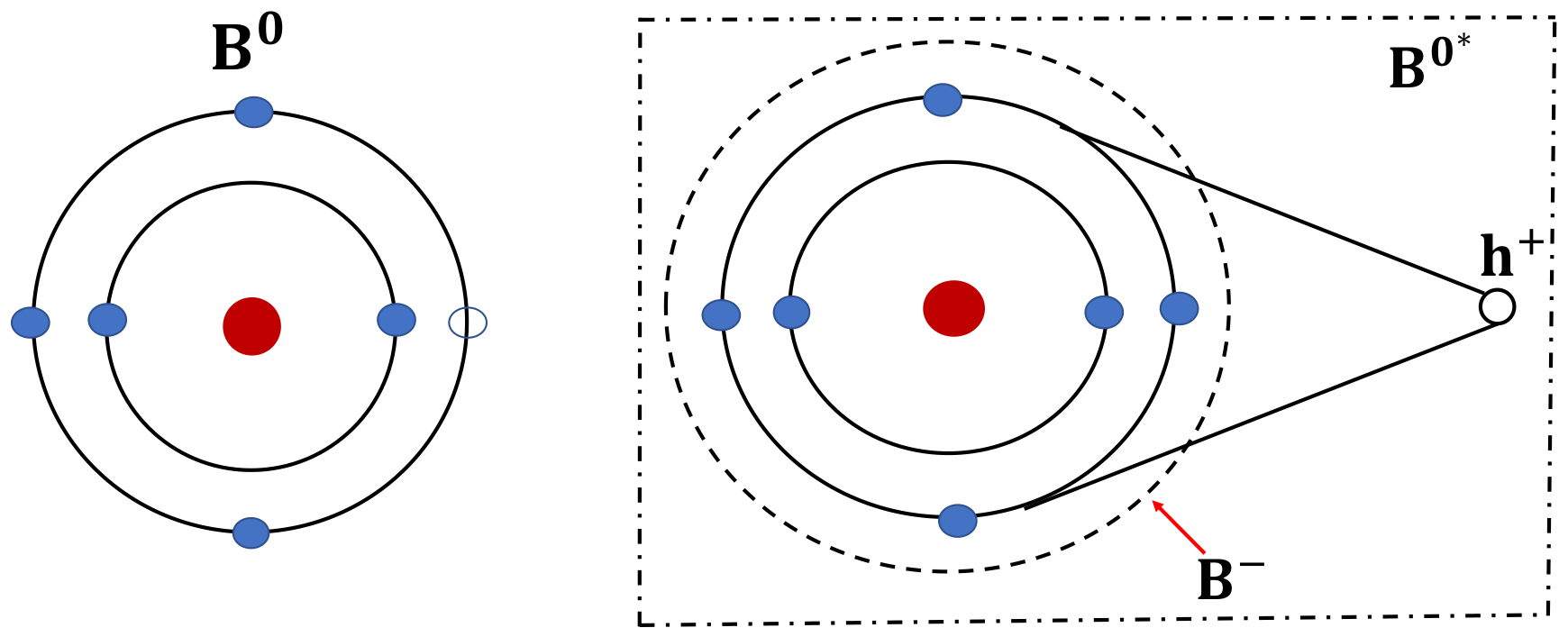}
  \caption{Displayed are the ground state of boron and a boron dipole state.}
  \label{fig:b}
\end{figure}

In the case of n-type impurities, phosphorus is prevalent in high-purity Ge detectors, making it a suitable example, as depicted in Figure~\ref{fig:p}. Conversely, for p-type Ge detectors, dominant impurities include boron, aluminum, and gallium, with boron chosen as an illustrative example, as shown in Figure~\ref{fig:b}. 

It's important to note that in their ground state, phosphorus and boron atoms are in their natural state. When incorporated into a Ge single crystal, these impurity atoms create covalent bonds with Ge atoms. Given that Ge possesses four electrons in its outermost orbital, phosphorus has five electrons in its outermost orbit, allowing the extra electron from phosphorus to become a free electron capable of moving independently from its parent ion. Similarly, the presence of free holes results from boron forming covalent bonds with Ge. However, when the temperature drops below 6.5 K, these free electrons and free holes become confined ("freeze-out") to their parent ions.
These bound states then stabilize as the temperature approaches 5.2 K, resulting in the formation of electric dipole states~\cite{mei2022evidence}. 

In summary, the capacitances measured without applied bias voltage from both n-type and p-type Ge detectors exhibit a rapid decrease from 11 K to 6.5 K. This phenomenon can be explained by the charge freeze-out theory, where free charges bound to donor states in n-type or acceptor states in p-type form electric dipole states.

\section{The Property of Electric Dipole States}
In this section, we initially introduce the concept that Ge detectors are anticipated to operate at temperatures below 6.5 K without depletion, as no free charges are expected to exist. Subsequently, we argue that this assumption may not hold for Ge detectors operating near helium temperatures. This is because the dipole states formed in the freeze-out process possess an electric field that can influence the drift field, consequently inducing charge trapping by forming cluster dipole states. We calculate the electric field strength for dipole states at three different temperatures, illustrating the temperature dependence of the field strength. Finally, we present the phenomenon of charge trapping by dipole states, leading to the formation of cluster dipole states.

\begin{figure} [htbp]
  \centering
  \includegraphics[clip,width=0.9\linewidth]{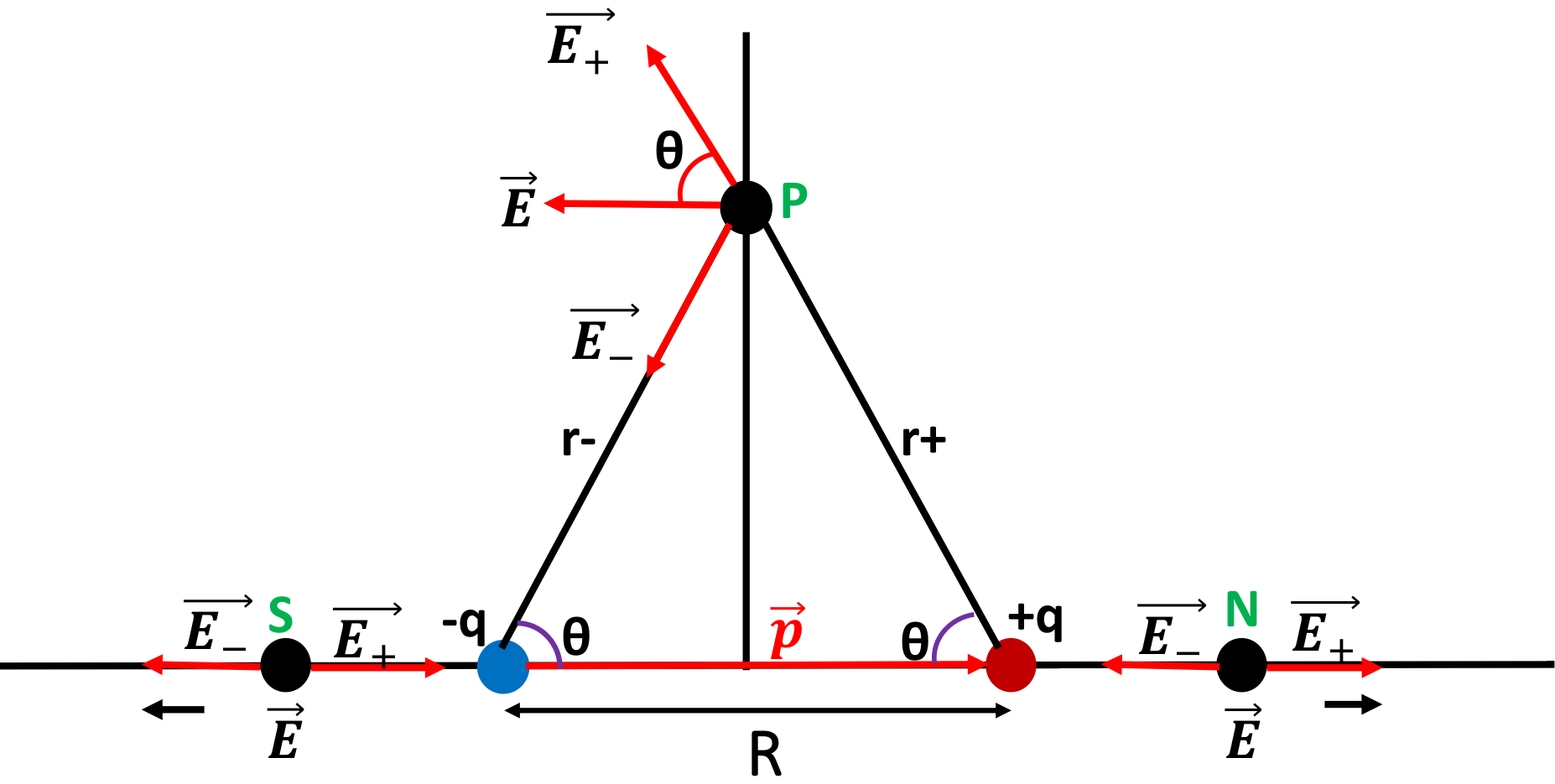}
  \caption{The displayed diagram depicts the field components at three positions in relation to the dipole ($\vec{p}$) created by a pair of charges. Point P lies along the perpendicular bisector, while Points S and N are positioned along the dipole's axis. Specifically, Point S is located near the charge $-q$ and Point N is close to the charge $+q$.}
  \label{fig:dipolefield}
\end{figure}

In general, when a Ge detector lacks free charge, it typically operates without the need for high-voltage depletion. This implies that Ge detectors can function at temperatures below 6.5 K with low voltages in the range of a few volts, a concept validated by experiments such as SuperCDMS~\cite{cdms} and EDELWEISS~\cite{ede}, where Ge detectors successfully operated at temperatures below 40 milliKelvin (mK). However, a different scenario emerged during the operation of planar detectors (R09-02 and RL) at 5.2 K at the University of South Dakota (USD). Both n-type (R09-02) and p-type (RL) detectors displayed notable charge trapping at low voltages below 100 volts~\cite{mei2022evidence}. The extent of trapping was contingent on the charge type drifting across the detector under the influence of an electric field. In an n-type detector, charge trapping was more pronounced when holes drifted across the detector, while in a p-type detector, electrons were more severely trapped during drifting~\cite{mei2022evidence}. This observation contrasts with Ge detectors operated below 40 mK in SuperCDMS and EDELWEISS, where Ge detectors performed well at voltages below 10 volts. Investigating the cause, we realized that the trapping of charges at lower electric fields can be attributed to the electric fields generated by dipole states. These electric fields diminish the overall field guiding drifting charges across the detector volume. Furthermore, the electric field strength produced by those dipole states must have a temperature dependence, which will be elucidated later. Figure~\ref{fig:dipolefield} illustrates the electric field generated by a dipole state comprising a pair of charges, one with a charge of $-q$ and the other with a charge of $+q$, at three distinct locations.

The electric field strength along the perpendicular bisector (Point P) can be calculated as follows:
\begin{equation}
    E = \frac{1}{4\pi\epsilon_0\epsilon}\frac{qR}{(r^2+(\frac{R}{2})^2)^{3/2}}.
\end{equation}
Here, $\epsilon_0$ represents the permittivity of free space, $\epsilon$ = 16.2 is the relative permittivity in Ge, $q$ denotes the unit of charge, $R$ signifies the distance between the two charges, the term $qR$ corresponds to the size of this dipole, and $r$ represents the distance between the center of the dipole and Point P. Similarly, the electric field strength along axis of dipole (Points S and N) can be expressed as:
\begin{equation}
    E = \frac{1}{4\pi\epsilon_0\epsilon}\frac{2rqR}{(r^2-{(\frac{R}{2})^2})^{2}},
\end{equation}
where $r$ represents the distance between the center of the dipole and either Point S or Point N, respectively.

\begin{figure} [htbp]
  \centering
  \includegraphics[clip,width=1.0\linewidth]{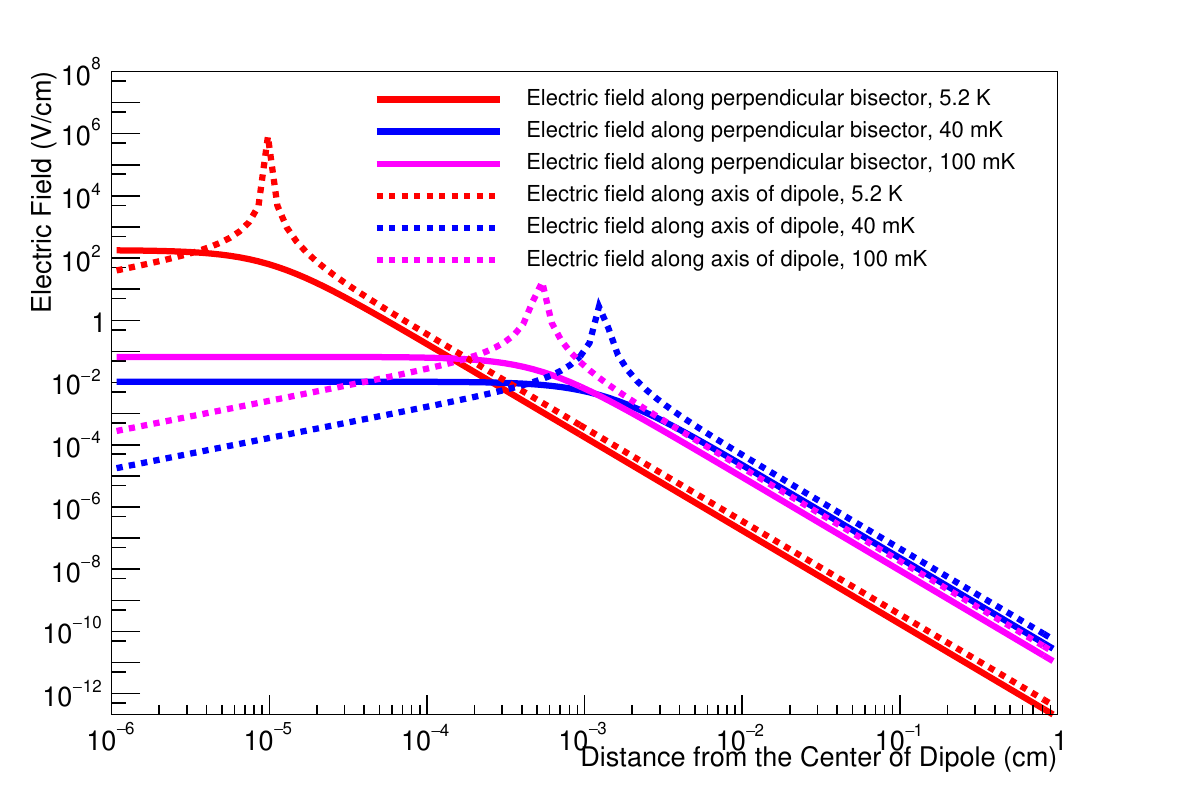}
  \caption{The graph displays the electric field strength as a function of the distance from the center of the dipole. Three temperatures are considered: one at 5.2 K, the second at 100 mK and the third at 40 mK.}
  \label{fig:fieldstrength}
\end{figure}

Note that the size of the dipole is highly dependent on temperature and is determined by the Onsager radius~\cite{onsager}, given by $r_c=\frac{e^2}{4\pi\epsilon\epsilon_{0} k_{B}T}$, in which charge carriers can be considered to be bound since
their mutual attraction energy $\frac{e^2}{4\pi\epsilon\epsilon_{0} r_{c}} > k_{B}T$,  where $e$ is the unit of electrical charge, $\epsilon=16.2$, and  $\epsilon_{0}$ is the permittivity of free space as defined before, $k_{B}$ is the Boltzmann constant, and $T$ denotes the temperature. This equation allows us to evaluate the strength of the electric field generated by a dipole. Figure~\ref{fig:fieldstrength} illustrates the electric field strength produced by dipole states as a function of the distance from the center of the dipole.

As illustrated in Figure~\ref{fig:fieldstrength}, the field strength generated by a dipole is notably higher at 5.2 K compared to 40 mK. This difference accounts for the Ge detector's ability to operate at low voltage, around a few volts, at 40 mK but not at 5.2 K. At 40 mK, the average field strength is below 10$^{-2}$ V/cm. The maximum field strength at 40 mK reaches approximately 10 V/cm for the field along the axis of the dipole and 10$^{-2}$ V/cm for the field along the perpendicular bisector. However, the peak feature of the field along the axis of the dipole occurs when the denominator in Equation (4) tends to zero. Contrastingly, at 5.2 K, within a few microns of distance, the average field strength is around 100 V/cm. This potent field significantly reduces the overall electric field for drifting charges across the detector, leading to charge trapping. Consequently, this reduction in the total field close to the dipole states results in the capture of charges, culminating in the formation of cluster dipole states.

\begin{figure} [htbp]
  \centering
  \includegraphics[clip,width=0.9\linewidth]{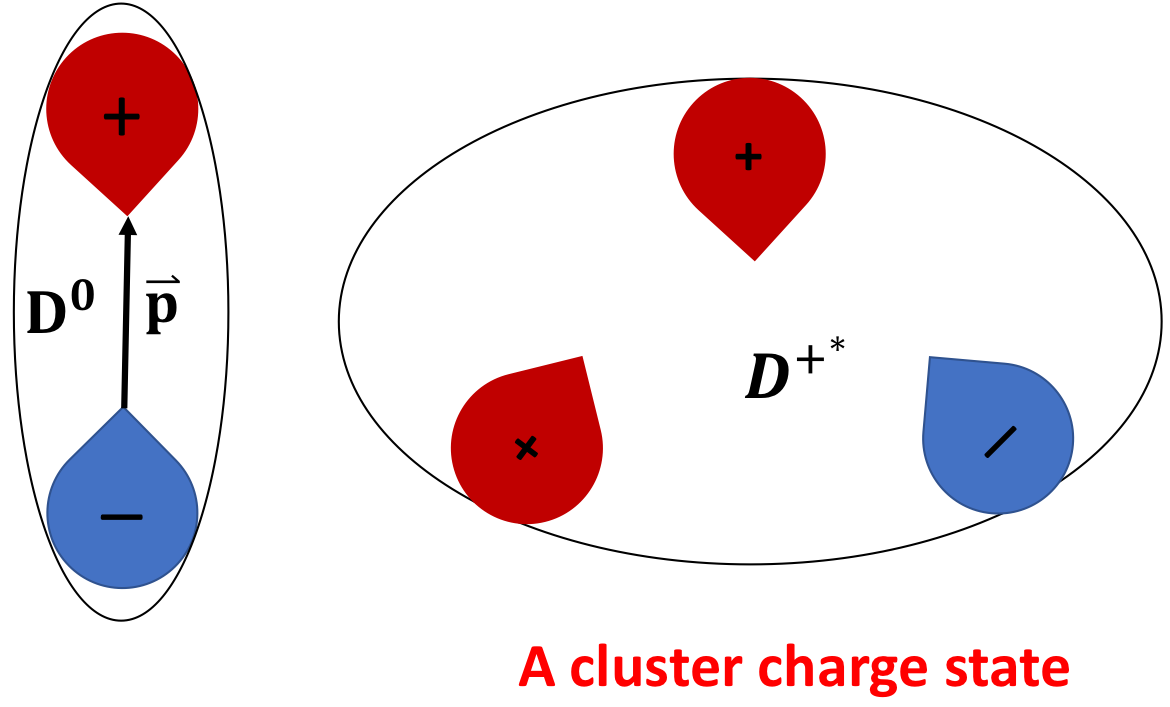}
  \caption{Presented are an n-type dipole state and an n-type cluster dipole state.}
  \label{fig:p1}
\end{figure}

\begin{figure} [htbp]
  \centering
  \includegraphics[clip,width=0.9\linewidth]{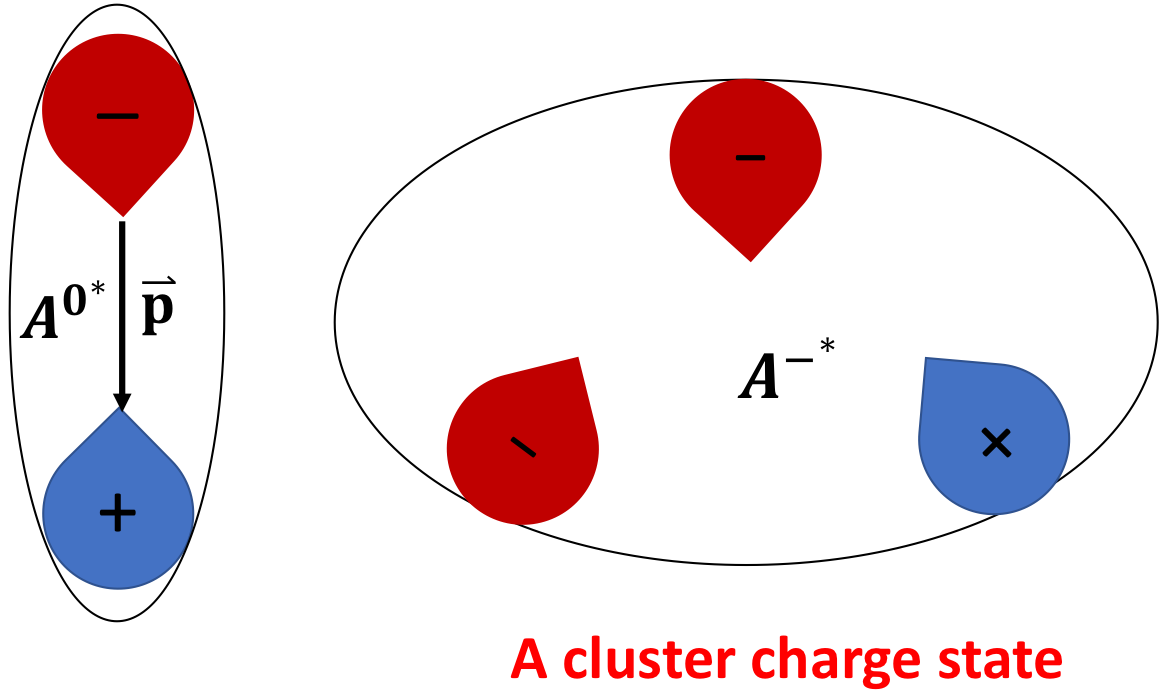}
  \caption{Presented are a p-type dipole state and a p-type cluster dipole state.}
  \label{fig:b1}
\end{figure}

In the case of an n-type Ge detector,  when an impurity atom is in its ground state, it has no capacity to trap charge. Upon cooling the detector to 5.2 K, a dipole state of an impurity atom forms, as illustrated in Figure~\ref{fig:p}. This n-type dipole state is capable of capturing charges, resulting in the formation of an n-type cluster dipole state, as demonstrated in Figure~\ref{fig:p1}.
Likewise, for a p-type Ge detector, when an impurity atom  is in its ground state, it cannot trap charges. However, once the detector is cooled down to 5.2 K, this impurity atom transition into a dipole state, becoming capable of trapping charges and forming a cluster dipole state. Figure~\ref{fig:b1} presents a p-type dipole state and a p-type cluster dipole state.

\begin{figure} [htbp]
  \centering
  \includegraphics[clip,width=0.9\linewidth]{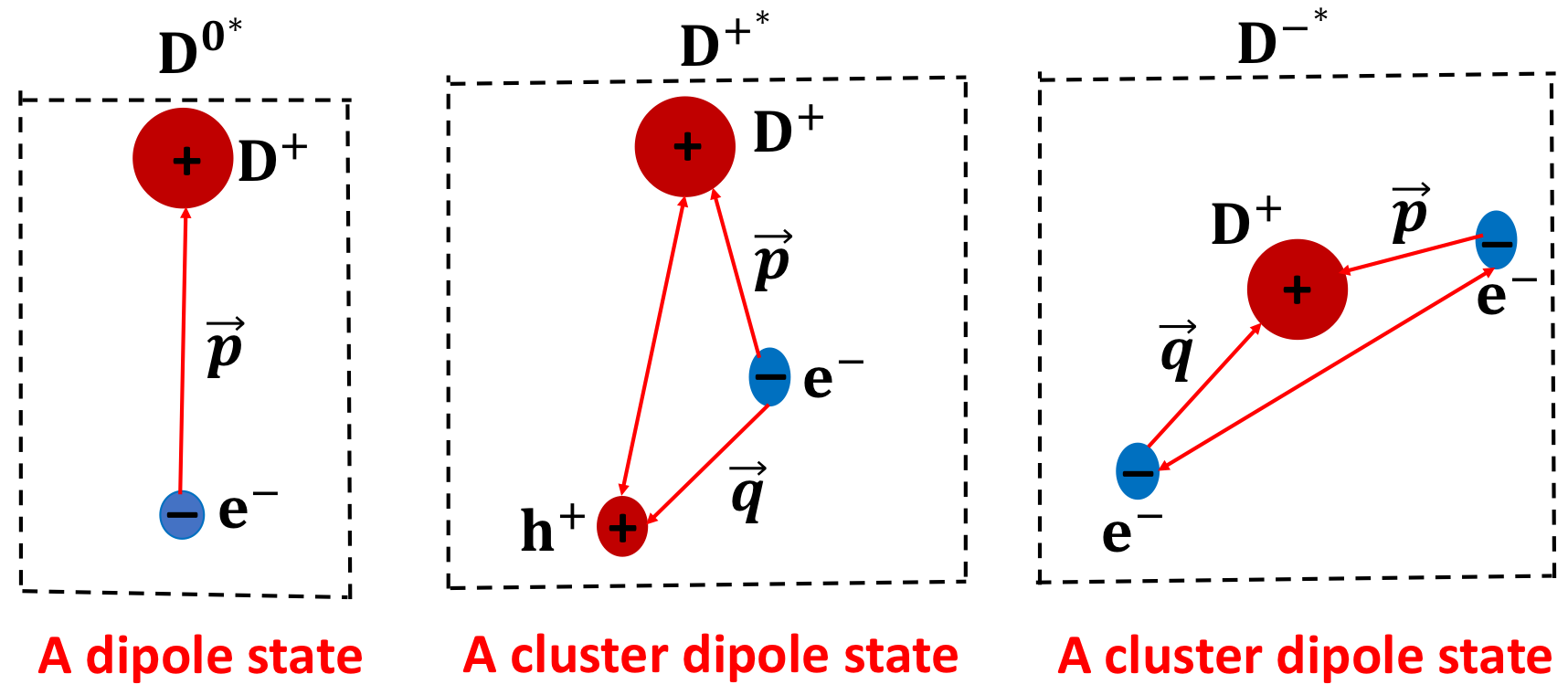}
  \caption{Shown are the processes involved in the formation of the dipole states and the cluster dipole states in an n-type Ge detector operated at 5.2 K, where $\vec{p}$ and $\vec{q}$ are the corresponding dipole moments~\cite{mei2022evidence}. $D^{0^{*}}$ stands for excited state that is differ from the ground state of $D^{0}$. }
  \label{fig:f00}
\end{figure}

 Figure~\ref{fig:f00} visually illustrates the formation of excited dipole states and cluster dipole states in n-type Ge, as discussed in our previous work~\cite{mei2022evidence}. Importantly, the probability for an immobile positive ion to trap charge carriers is relatively smaller than that for movable bound electrons. The movable space for electrons is approximately equal to a sphere formed by the Onsager radius stated earlier, within which charges can be trapped, while an immobile positive ion traps charges only when they come close to it. Consequently, the likelihood of forming $D^{+^{*}}$ states is higher compared to the formation of $D^{-^{*}}$ states in an n-type detector. Note that $D^{+^{*}}$ and $D^{-^{*}}$ are excited states that differ from the ground states of $D^{+}$ and $D^{-}$. This is the key reason why holes become trapped more extensively than electrons in an n-type detector.

\begin{figure} [htbp]
  \centering
  \includegraphics[clip,width=0.9\linewidth]{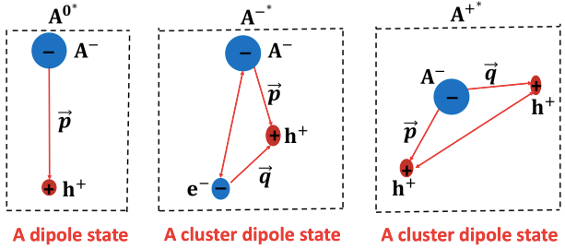}
  \caption{Shown are the processes involved in the formation of the excited dipole states and the cluster dipole states in a p-type Ge detector operated at low temperatures, where $\vec{p}$ and $\vec{q}$ are the corresponding dipole moments~\cite{mei2022evidence}. $A^{0^{*}}$ represents excited state which is differ from the ground state of $A^{0}$. }
  \label{fig:f0}
\end{figure}

 The generation of excited dipole states and cluster dipole states in p-type Ge is depicted in Figure~\ref{fig:f0}~\cite{mei2022evidence}. The available space for immobilized negative ions, which trap charge carriers, is smaller than that for mobile bound holes. The movement of holes is defined by the Onsager radius, denoted as $r_{c}$ as defined earlier. Since the trapping probability is proportional to the available space where charges can be trapped, the probability of trapping for immobilized negative ions is smaller than that of movable holes. Consequently, the likelihood of creating $A^{-^{*}}$ states is greater than that of forming $A^{+^{*}}$ states in a p-type detector. Similarly, $A^{+^{*}}$ and $A^{-^{*}}$ are excited states that differ from the ground states of $A^{+}$ and $A^{-}$.This explains why electrons are more severely trapped than holes in a p-type detector.

These formed dipole states and their associated cluster dipole states demonstrate remarkable thermal stability at 5.2 K. Previous in-depth studies~\cite{san,math} have meticulously scrutinized the binding energies of these states in both n-type and p-type detectors. Their consistent findings reveal binding energy values falling within the range of 5-8 meV, for both n-type and p-type impurities~\cite{san,math}. This low binding energy characteristic suggests that Ge detectors with residual impurities at 5.2 K hold significant potential for applications in the search for MeV-scale DM.

In summary, our investigation delved into the attributes of electric dipole states produced during the freeze-out process in both n-type and p-type Ge detectors. We observed a temperature-dependent variation in the size of these dipole states, which consequently impacts the strength of the associated electric field. Importantly, these dipole states exhibit the capacity to capture drifting charges, giving rise to the formation of cluster dipole states. Notably, at the temperatures around 5.2 K, these cluster dipole states demonstrate thermal stability~\cite{san,math}.

 \section{Design a planar Ge detector for GeICA implementation.}
 In this section, we introduce the concept of a Ge detector designed for operation at 5.2 K, specifically tailored for the direct detection of MeV-scale DM. We delve into the impact ionization of dipole states, deriving the amplification factor as a function of impurity levels. Our analysis leads to the conclusion that an amplification factor of 100 can be achieved with an impurity level of 2$\times$10$^{11}$/cm$^3$.
 
 USD hosts a distinctive Ge detector production facility, encompassing Ge zone refining~\cite{yang1,yang2,yang3}, crystal growth~\cite{wang1,wang2,wang3}, crystal characterization~\cite{raut}, and detector fabrication~\cite{meng, wei, panth}. Moreover, USD has pioneered the development of planar Ge detectors~\cite{meng, wei, panth} employing locally cultivated crystals~\cite{wang1, wang2, wang3}. Leveraging this capability, a planar p-type Ge detector, measuring 7 cm × 7 cm × 4 cm, can be manufactured for optimal performance at 5.2 K. This Ge detector boasts a total mass of 1.04 kg. The contacts can be constructed using amorphous Ge, following a design akin to the previously established planar detectors at USD~\cite{meng}. Upon cooling the detector to 5.2 K, it will give rise to the formation of excited A$^{0^{*}}$ states. The binding energies of the excited dipole states and cluster dipole states at 5.2 K were examined for both n-type and p-type detectors, as documented in previous studies~\cite{san, math}. These investigations consistently reported binding energy values in the range of 5-8 meV, for both n-type and p-type impurities~\cite{san, math}.

Consider a MeV-scale dark matter particle interacting with a Ge atom, depositing a mere 0.1 eV of energy. Such a modest energy deposition initiates phonon emission, with individual phonons carrying energy around 0.04 eV, as calculated by Mei et al.\cite{mei}. The transportation of these phonons is also elucidated by Mei et al.\cite{mei}. When these phonons are absorbed by an excited dipole state, A$^{0^{*}}$, they release a free hole charge. Under the influence of a strong electric field that can overcome trapping caused by the formation of cluster dipole states, a free hole is able to drift across the detector without being trapped. During this progression, the drifting hole gains kinetic energy through the work done by the electric field, expressed as $eEl$, where $e$ signifies the electric charge, $E$ denotes the strength of the electric field, and $l$ represents the drift length along the field direction. In the presence of a sufficiently robust electric field, the drifting hole amasses enough kinetic energy to surpass 0.01 eV along its drift path, potentially leading to the impact ionization of excited dipole states along its trajectory toward the electrode, where charges are ultimately collected. This impact ionization process generates additional holes through $h$ + A$^{0^{*}}$ $\rightarrow$ 2$h$ + $A^{-}$, thereby amplifying the charge signal. The increase in the number of charge carriers per unit of time is defined by dividing the total number of charges generated in this process by the generation time. This relationship can be expressed as follows:
\begin{equation}
    \frac{dn}{dt} = \frac{n}{\tau_{imp}},
\end{equation}
where $\tau_{\text{imp}}$ represents the generation time, and $n$ signifies the number of charge carriers generated during this time interval. Note that we have defined the time derivative as a positive value to signify the generation of carriers through impact ionization. Solving this equation yields the following:
\begin{equation}
    n(t) = n_{0}exp(\frac{t}{\tau_{imp}}). 
\end{equation}
The amplification factor, denoted as $A$, is defined by the formula $A = \frac{n(t)}{n_{0}}$. In this equation, $t$ represents the drift time, calculated as $t = \frac{d}{v_{d}}$, where $d$ corresponds to the drift length, and $v_{d}$ is the drift velocity. The generation time $\tau_{imp}$ can be expressed as $\tau_{imp} = \frac{1}{N_{d}\sigma(E) v_{d}}$. Here, $N_d$ stands for the impurity density, and $\sigma(E)$ is approximately equal to $\sigma_0 \times \sqrt{E}$~\cite{umn}, denoting the cross-section for impact ionization of impurities, which varies with the applied electric field. The constant $\sigma_0$ is defined as $\sigma_0$ = 5$\times10^{-13}$ cm$^2$~\cite{phipps}. Combining all of these factors, the amplification factor can be expressed as:
\begin{equation}
    A = exp(N_{d}\sigma(E) d).
\end{equation}

\begin{figure} [htbp]
  \centering
  \includegraphics[clip,width=0.9\linewidth]{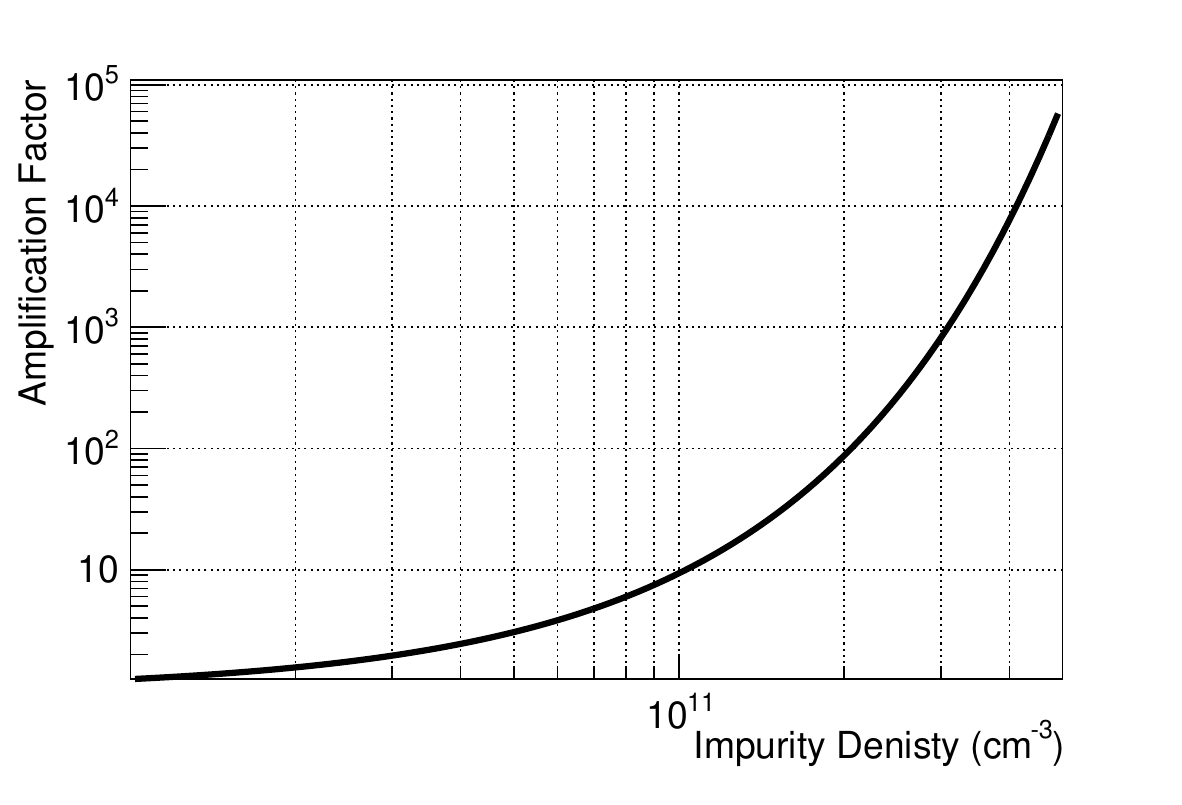}
  \caption{The predicted amplification is illustrated as a function of impurity concentration. This prediction is based on an assumed average drift length of 2 cm within the detector and an electric field of 500 V/cm.}
  \label{fig:amp}
\end{figure}

The extent of amplification is contingent on the impurity concentration in the detector, the impact ionization cross-section, and the drift length, which correlates with the detector's thickness. Assuming an average drift length of 2 cm within the detector and an average electric field of 500 V/cm, the amplification factor concerning impurity concentration can be visualized in Figure~\ref{fig:amp}. It is apparent that an amplification factor of 100 can be attained at an impurity level of 2$\times$10$^{11}$/cm$^{3}$. 
Our previous studies indicate that an amplification factor of 100 can be achieved using a similar detector at 40 mK~\cite{umn}. 
 
In summary, we introduce the concept of a Ge detector tailored for direct DM detection, leveraging populated dipole states at 5.2 K to achieve an amplification factor of 100 and attain a detection energy threshold of approximately 0.1 eV for energy deposited by DM interacting with ordinary matter. Our analysis demonstrates that with an impurity level of 2$\times$10$^{11}$/cm$^3$ and operation at 5.2 K with a biased field of 500 V/cm, this proposed detector can indeed achieve the targeted amplification factor of 100.

\section{Conclusion}
In conclusion, our investigation of Ge properties at helium temperatures has yielded a fascinating revelation—a significant decrease in measured relative capacitance from 11 K to 6.5 K. This change is attributed to a remarkable phenomenon, where free charges within the Ge detector bound to their parent ions within this specific temperature range. As the temperature drops, these impurity atoms, previously in free charge states, undergo a transition into stable localized dipole states. These dipole states, characterized by a binding energy within the range of 5 - 8 meV, exhibit an intriguing level of thermal stability.

What's particularly noteworthy is the behavior of these dipole states when subjected to phonon absorption. They have the capacity to undergo ionization under these conditions, resulting in the liberation of free charges within the detector volume. What's more, by applying an appropriate bias voltage to the detector, these charges can be efficiently drifted. As they drift, they gain kinetic energy, culminating in the generation of additional charges through the impact ionization of dipole states. This intricate impact ionization process plays a pivotal role as a signal amplification mechanism within the Ge detector. In practical terms, it efficiently amplifies the charge signals, allowing them to significantly surpass the levels of electronic noise typically encountered in such detectors.

This phenomenon carries significant implications, especially in scenarios where MeV-scale DM interacts with ordinary matter. These interactions predominantly involve the emission of phonons, and the signal amplification achieved through the impact ionization of dipole states can be invaluable in effectively detecting and distinguishing these faint interactions.

As a direct outcome of these compelling findings, we foresee the development of a Ge detector with an estimated mass of approximately 1 kg. Such a detector has the potential to achieve an impressively low detection threshold of less than 0.1 eV. To reach this threshold, an amplification factor of 100 can be realized with an impurity level of 2$\times$10$^{11}$/cm$^3$ under an operating field of 500 V/cm.
Please bear in mind that when an energy deposition of 0.1 eV occurs, it can generate at least a single electron-hole pair through the ionization of excited dipole states with a binding energy of 0.01 eV. With an amplification factor of 100, this translates to a total of 100 charge carriers contributing to the detection signal. To provide context, this scenario resembles the ionization of Ge atoms in a conventional Ge detector. In such detectors, a single charge carrier is typically generated with an average energy deposition of 3 eV~\cite{Wei02}, leading to an accumulated energy deposition of 300 eV to yield 100 charge carriers. The current detection energy threshold for a conventional Ge detector is limited by the intrinsic electronic noise, typically around $\sim$100 eV~\cite{Barton}. The proposed detector, utilizing impact ionization of impurities with an amplification factor of 100, can achieve a detection threshold as low as 0.1 eV. 

In practical terms, harnessing impurity ionization holds the potential for a substantial reduction in detection energy thresholds. This transformative shift departs from the conventional Ge detector, requiring 100 eV to overcome electronic noise, to an impressive threshold of 0.1 eV. This lowered threshold is further magnified by a factor of 100, surpassing electronic noise. Such an enhancement represents a remarkable improvement by a factor of 1000, enabling exploration across a wide parameter space ranging from the GeV/c$^2$ down to below the MeV/c$^2$ range in the pursuit of DM investigation. 

It is important to underscore that the leakage current represents a potential technical impediment to achieving the desired low energy detection threshold. In planar detectors or planar point-contact detectors fabricated at USD, the leakage current arises from two principal sources: injection leakage current, stemming from the reduced barrier height of amorphous contacts at higher fields, and surface leakage current, linked to the heightened hopping effect at elevated fields. The application of higher fields, such as 500 V/cm, may lead to an increase in both injection and surface leakage currents owing to field penetration.

To mitigate surface leakage current, the implementation of a guarding ring, as demonstrated by Wei et al.\cite{weimei}, stands out as a notable strategy. The potential elevation of noise levels and its impact on the low-energy threshold due to injection leakage from the contacts underscores the significance of addressing these challenges. Currently, in collaboration with Pacific Northwest National Laboratory, we are conducting tests on a similar detector\cite{weimei} within a low-noise system. The outcome of this research is poised to provide deeper insights into these phenomena. If validated, this pioneering advancement will open new horizons in direct DM detection, representing a significant leap forward in our pursuit to unravel the mysteries of the universe.
 
 \section{Acknowledgement} 
 We extend our gratitude to Kyler Kooi, Rajendra Panth, and Jing Liu for their contributions to the construction of the cryostat designed for use with the Pulse Tube Refrigerator. Special acknowledgment goes to Kyler Kooi, Rajendra Panth, Sanjay Bhattarai, and Mathbar Raut for their significant roles in conducting relative capacitance measurements, as previously detailed in our published work. This work was supported in part by NSF OISE 1743790, NSF PHYS 2310027, DOE DE-SC0024519, DE-SC0004768, DE-SC0024519, and a research center supported by the State of South Dakota.

\end{document}